\documentclass[prl,twocolumn,aps,superscriptaddress,showpacs]{revtex4}
\usepackage{graphicx}
\usepackage{color}
\usepackage{amsmath}
\usepackage{mathrsfs}
\usepackage{dsfont}
\usepackage{amstext}
\usepackage{amssymb}
\usepackage{amsbsy}
\usepackage{bbm}
\usepackage{amsthm}

\begin{document}
\newcommand{\pprl}{Phys. Rev. Lett. \ }
\newcommand{\pprb}{Phys. Rev. {B} }
\newcommand{\tphi}{{\tilde\phi}}
\newcommand{\tb}[1]{{\textbf{#1}}}
\newcommand{\dtwo}[1]{{\text{d}^2\textbf{#1}}}
\newcommand{\gl}[1]{{g^{(l)}_{{#1}}}}
\newcommand{\tgl}[1]{{\tilde g^{(l)}_{{#1}}}}
\newcommand{\fl}[1]{{f^{(l)}({#1})}}
\newcommand{\Rp}[1]{{\frac{R_{#1}}{1+R_{#1}\tilde R_{#1}}}}
\newcommand{\tRp}[1]{{\frac{\tilde R_{#1}}{1+R_{#1}\tilde R_{#1}}}}
\newcommand{\tR}[1]{{\tilde R}_{#1}}
\newcommand{\tD}[1]{{\tilde\Delta}_{#1}}
\newcommand{\flt}[1]{{f^{(#1)}}}
\newcommand{\imth}{\hspace{1pt}\mathrm{i}\hspace{1pt}}
\newcommand{\ie}{{\it i.e.~}}

\title{Correlation-hole induced paired quantum Hall states in lowest Landau level}

\author{Yuan-Ming Lu}
\affiliation{Institute of Theoretical Physics, Chinese Academy of
Sciences, P.O. Box 2735, Beijing 100190, China}
\affiliation{Department of Physics, Boston College, Chestnut Hill,
MA 02467}
\author{Yue Yu} \affiliation{Institute of Theoretical Physics, Chinese Academy of
Sciences, P.O. Box 2735, Beijing 100190, China}
\author{Ziqiang Wang}
\affiliation{Department of Physics, Boston College, Chestnut Hill,
MA 02467} \affiliation{Institute of Theoretical Physics, Chinese
Academy of Sciences, P.O. Box 2735, Beijing 100190, China}

\date{\today}

\begin{abstract}
A theory is developed for the paired even-denominator fractional
quantum Hall states in the lowest Landau level. We show that
electrons bind to quantized vortices to form composite fermions,
interacting through an exact instantaneous interaction that favors
chiral $p$-wave pairing. There are two canonically dual pairing gap
functions related by the bosonic Laughlin wavefunction (Jastrow
factor) due to the correlation holes. We find that the ground state
is the Moore-Read pfaffian in the long wavelength limit for weak
Coulomb interactions, a new pfaffian with an oscillatory pairing
function for intermediate interactions, and a Read-Rezayi composite
Fermi liquid beyond a critical interaction strength. Our findings
are consistent with recent experimental observations of the $1/2$
and $1/4$ fractional quantum Hall effects in asymmetric wide quantum
wells.
\end{abstract}


\maketitle

\setcounter{tocdepth}{3}

The fractional quantum Hall effect (FQHE) observed at Landau level
filling fraction $\nu=5/2$ \cite{will} signifies a new state of
correlated electrons. This state is believed to be described by the Moore-Read pfaffian (MRP) \cite{MR} and supports fractionalized quasiparticle
excitations \cite{heiblummarcus} that obey nonabelian statistics relevant for topological quantum computing \cite{kitaev}. An outstanding question has been
whether such nonabelian topological phases exist
in the lowest Landau level (LLL). Several recent experiments
\cite{exp1,exp2,exp3} indeed observed FQHE at $\nu=1/2$ and $1/4$,
suggesting that these, too, may be in the MRP phase. Although
the abelian two-component Halperin (331) and (553) states
\cite{halp} can be strong contenders for these FQHE \cite{papic1},
fresh experiments and numerical studies found strong evidence for the
one-component FQHE at $\nu=1/2$ and $1/4$ in
asymmetric wide quantum wells \cite{exp3,papic2}. Whether the observed FQHE
can be understood as pfaffians in the LLL is the focus
of this work.

The MRP is 
a chiral $p$-wave paired quantum Hall state \cite{readgreen}. In
principle, it can emerge as a $p$-wave pairing instability of the
composite Fermi liquid (CFL), a gapless state of electrons attached
to flux tubes \cite{HLR}. The leading-order statistical interaction
mediated by the Chern-Simons (CS) gauge field fluctuations can
produce a $p$-wave pairing potential for the composite fermion
\cite{GWZ} (CF). However, since the coupling between the CF and the
CS gauge field is not small, diagrammatic perturbation theory is not
controllable. Within the random-phase approximation, the gauge
fluctuations are in fact singular and pair-breaking \cite{nb}.
Therefore, the ground states at filling fractions $1/2$ and $1/4$
remained enigmatic \cite{sankar}.

The key to solve this problem is to properly account for the effects
of the correlation hole, i.e. the local charge depletion caused by
attaching flux to an electron. A CF feels the correlation hole of the other CFs,
which is captured by the Jastrow factor in the
Laughlin wavefunction. In the unitary CF theory \cite{jain,HLR},
only an infinitely thin flux tube associated with a U(1) phase is attached to each
electron without accounting for the Jastrow factor, i.e. the
correlation hole. Read improved the concept of
CF by attaching finite size vortices
to electrons \cite{nread,ReR} such that the Jastrow factor naturally appears in the ground state wavefunction. Binding zeros of the wavefunction to electrons keeps
them apart and lowers the Coulomb energy. The vortex attachment can
be achieved through a non-unitary but nevertheless canonical
transformation on the electron operators \cite{RS}. The saddle point
solutions of such a non-unitary CF (NUCF) field theory recover the
Laughlin state for the odd denominator FQHE \cite{RS} and the
Rezayi-Read CFL at $\nu=1/2$ \cite{yw}. The effective interaction induced by the vortex attachment has also been studied at $\nu=1/2$ \cite{japan}.

In this paper, we show that paired quantum Hall states emerge in the
LLL using the NUCF field theory where a vortex with
vorticity-$\tphi$ ($\tphi=$ even integer) is attached to an electron
at filling fraction $\nu=1/\tphi$. An important feature
of attaching vortices to electrons is that the diamagnetic coupling,
quadratic in the gauge field, to the CF density is canceled by its
dual contribution from the correlation hole associated with the
vortex. As a result, we show that the gauge fluctuations can be
integrated out exactly, leading to an {\em instantaneous}
statistical interaction between the NUCFs which is attractive for
chiral $p_x-\imth p_y$ pairing \cite{japan} and scales linearly with
the vorticity $\tphi$.
We construct the variational ground state
wavefunction for such a {\it correlation-hole induced paired quantum
Hall state}, introducing two canonically dual gap functions related
by the radial distribution function of the corresponding bosonic
Laughlin wavefunction. Variational calculations are carried out in
the presence of the pair-breaking Coulomb interaction
$V_c(r)=e^2/4\pi\epsilon_r\epsilon_0r=\lambda\frac{2\sqrt{2\nu}\ell_B}{r}\hbar\omega_c$,
where $\ell_B$ is the magnetic length and
$\lambda$ is a dimensionless coupling constant. We find that for
weak Coulomb interaction $\lambda <\lambda_{c1}$, the ground state
is a MRP in the long wavelength limit. However, the pairing function
deviates significantly from that of the MRP at shorter distances,
consistent with recent numeric studies in the projected LLL
\cite{moller-simon}. Remarkably, we find that for intermediate
Coulomb interactions, $\lambda_{c1} < \lambda < \lambda_{c2}$, the
paired state is different from the MRP even in the asymptotic long
wavelength limit. The pairing function in this new phase is
oscillatory with its amplitude decaying as the inverse square root
of the distance. For sufficiently strong Coulomb interactions
$\lambda>\lambda_{c2}$, the paired state becomes unstable and the
ground state undergoes a transition to the Rezayi-Read CFL\cite{ReR}
phase. The topological properties and the
effect of a short-ranged interaction are also studied.

We consider 2D spin-polarized interacting electrons described by the
field operator $\psi_e$ in a uniform perpendicular magnetic field
$B$. The electron density is $n_e$ and the density operator
$\rho=\psi_e^\dagger\psi_e$. The vortex attachment is
through the following non-unitary transformation \cite{RS}:
\begin{eqnarray}\label{nutCF}
\Phi({\bf r})=e^{-J_{\tilde\phi}({\bf r})}\psi_e({\bf r}),~~~
\Pi({\bf r})=\psi_e^\dagger({\bf r})e^{J_{\tilde\phi}({\bf r})},
\end{eqnarray}
where $J_{\tilde\phi}({\bf r})=I_{\tilde\phi}({\bf
r})-\frac{|z|^2}{4l^2_ {\tilde\phi}}$,
$\ell_{\tilde\phi}=\sqrt{\frac{\hbar c}{\nu\tilde\phi eB}}$, and
\begin{eqnarray}
&I_{\tilde\phi}({\bf r})={\tilde\phi}\int d^2{\bf r}'\rho({\bf
r}')\log(z-z'), \quad z=x+iy. \nonumber
\end{eqnarray}
We set $\hbar=|e|/c=1$ hereafter. The imaginary part of $I_\tphi$
coincides with the unitary CS transformation, while its real part
describes the finite vortex core (correlation hole) accompanying the
flux attachment. Note that the fields $\Phi$ and $\Pi$ are not
hermitian conjugate, $ \Pi=\Phi^\dagger
e^{J_{\tilde\phi}+J^\dagger_{\tilde\phi}}$. However, they form a
pair of canonical fields obeying fermion anti-commutation relations;
the operator $\Pi$ creates a NUCF while $\Phi$ annihilates one and
$\rho=\Pi\Phi$. The transformed Hamiltonian reads
\begin{eqnarray}
&\nonumber H^{\rm CF}=\frac{1}{2m_b}\int d^2{\bf r} \Pi({\bf
r})[-\imth\nabla-({\bf A} -{\bf v}_{\tilde\phi})]^2
\Phi({\bf r}) \nonumber\\
&+\frac{1}{2}\int d^2{\bf r} d^2{\bf r}'\delta\rho({\bf r})V_c({\bf
r}-{\bf r}') \delta\rho({\bf r}') \label{CFH}
\end{eqnarray}
where ${\bf v}_{\tilde\phi}({\bf r})={-}\imth\nabla J_{\tilde\phi}=
{\bf a}(\vec{x}){+}\imth\hat{n}\times{\bf a} ({\bf
r})-\imth\frac{{B}}{2}{\bf r} $ with $\hat n$ normal to the 2D plane
and the CS gauge field is
\begin{equation}
{\bf a}({\bf r})={\tilde\phi}\nabla\int d^2{\bf r}'\rho({\bf r}')
{\rm Im}\log(z-z'). \label{A}
\end{equation}
The statistical magnetic field $ b=\nabla\times {\bf
a}=2\pi{\tilde\phi}\rho$.

One of the physical justifications to introduce such a NUCF field
theory lies in the fact that the resulting mean-field states give
rise to numerically well-tested wavefunctions \cite{RS,yw}. At the
mean field level, one takes gauge field ${\bf a}$ to be a classical
one determined by (\ref{A}) with a uniform density $\bar{\rho}({\bf
r})=n_{e}$, and $ \bar{\bf a}({\bf r})={-}\frac{\nu\tilde\phi
B}{2}\hat n\times {\bf r}$. Thus, the mean-field theory describes
free NUCFs in an effective magnetic field $\Delta B=B-\nu\tilde\phi
B=\nabla\times \Delta {\bf A}$ with $\Delta {\bf A}={\bf A}-\bar{\bf
a}$. At even-denominator filling fractions $\nu=1/\tphi$,
$\ell_\tphi=\ell_B$ and the effective $\Delta B=0$, the mean-field
ground state is the Rezayi-Read CFL \cite{ReR}.

An important, non-perturbative feature of this NUCF theory is that
the usual diamagnetic fermion density-gauge field coupling of the
form $\rho\delta{\bf a}^2$, where $\delta{\bf a}={\bf a}-\bar{\bf
a}$, is canceled in Eq.~(\ref{CFH}) by the contribution from the
correlation holes since $(\delta{\bf a}\pm\imth\hat
n\times\delta{\bf a})^2=0$. As a result, the gauge fluctuations in
the CS action can be exactly integrated out to obtain a closed-form
effective Hamiltonian:
\begin{eqnarray}\label{HeffMomentum}
H_{\rm eff}^{\rm CF}&=&\sum_\textbf{k}
(\xi_\textbf{k}+\frac{\omega_c}{2})\Pi_\textbf{k}\Phi_\textbf{k} \nonumber\\
&+&\frac{\pi\tilde\phi}{m_b}\sum_{\textbf{q}\neq0,\textbf{k},\textbf{p}}\frac{k+p}{q}
\Pi_{\textbf{k}-\textbf{q}}\Pi_{\textbf{q}-\textbf{p}}
\Phi_{-\textbf{p}}\Phi_\textbf{k} \nonumber \\
&+&\sum_{\textbf{q}\neq0,\textbf{k},\textbf{p}}\frac{v_\textbf{q}}{2}
\Pi_{\textbf{k}-\textbf{q}}\Phi_\textbf{k}\Pi_{\textbf{q}-\textbf{p}}
\Phi_{-\textbf{p}}\Phi_{-\textbf{p}}\Phi_\textbf{k}.
\end{eqnarray}
Here $\xi_\textbf{k}=\frac{\hbar^2|\textbf{k}|^2}{2m_b}-\mu$ and
$\omega_c=|B|/m_b$ is the cyclotron frequency. The second term is
the induced {\em instantaneous} statistical interaction written in
terms of the complex momenta $k=k_x+\imth k_y$ (similarly for $p$,
$q$). For $\textbf{k}=\textbf{p}$, it reduces to a singular pairing
interaction that scatters a pair of NUCFs with zero center-of-mass
momentum from $(\textbf{k},-\textbf{k})$ to
$(\textbf{k}^\prime,-\textbf{k}^\prime)$ with momentum transfer
$\textbf{q}=\textbf{k}-\textbf{k}^\prime$. Expanding in the
angular-momentum channels ($l$):
\begin{eqnarray}\label{StatisticalAttraction}
\frac{1}{2}\frac{k+p}{k-k^\prime}\Big
|_{p=k}=1+{\sum_{l\neq0}\rm
sign}(l)\left\vert\frac{k^\prime}k\right\vert^le^{\imth
l\theta_{kk^\prime}}\theta(1-\left\vert\frac{k^\prime}k\right\vert^l)
\nonumber
\end{eqnarray}
where $\theta_{kk^\prime}=\arg(k^\prime)-\arg(k)$, we see that the
pairing potential is attractive for $l<0$ 
with dominant $p_x-\imth p_y$ scattering. The Coulomb interaction in
Eq.~(\ref{HeffMomentum}), where
$v_\textbf{q}={e^2}/{2\varepsilon_r\varepsilon_0|\textbf{q}|}$, is
purely repulsive in all channels. In the absence of Coulomb
interaction, it can be shown that the MRP, being an antiholomorphic
function, is an exact ground state of the NUCF Hamiltonian.

\setlength{\unitlength}{1in}
\begin{figure}
\begin{picture}(3.7,3.2)
\put(-0.3,0){\makebox(3.7,3.2){\includegraphics[width=3.7in]{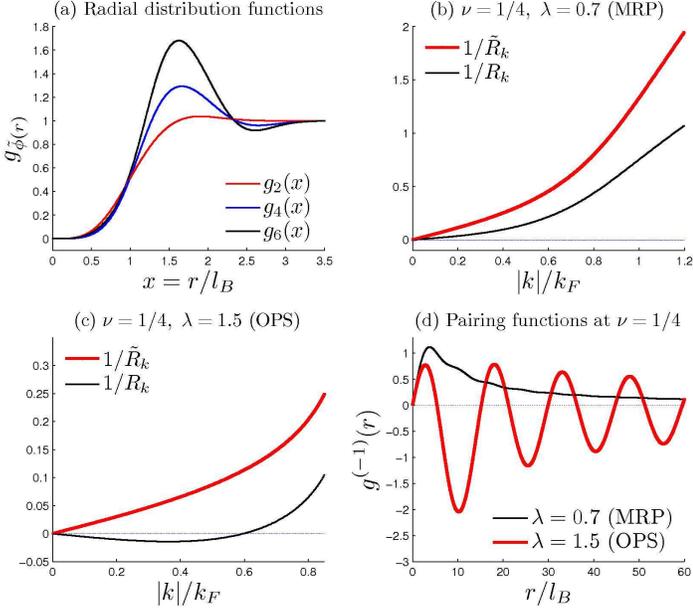}}}
\end{picture}
\caption{(a): Radial distribution function $g_\tphi$ with even
$\tphi=1/\nu$ from Monte-Carlo \cite{ciftja}. (b) and (c): Inverse
pairing functions $1/R_k$ and $1/\tilde R_k$ at $\nu=1/4$ for two
different Coulomb strengths $\lambda$ in the MRP phase (b) and the
OPS phase (c). (d) Real-space chiral $p$-wave pairing function
$g^{(-1)}(r)$ in the two phases (b) and (c). At $\nu=1/4$,
$\lambda_{c1}\approx1.0$.} \label{fig1}
\end{figure}

To study the variational ground state of Hamiltonian
(\ref{HeffMomentum}), we generalize the BCS theory to the
non-unitary case. Introduce the Dirac ket and bra 
\begin{eqnarray}\label{PairedState}
|G^{(l)}\rangle\equiv
e^{\frac{1}{2}\sum_\textbf{k}g_\textbf{k}^{(l)}\Pi_\textbf{k}
\Pi_{-\textbf{k}}}|0\rangle,~~\langle
G^{(l)}|=\langle0|e^{\frac{1}{2}\sum_\textbf{k}\tilde
g_\textbf{k}^{(l)}\Phi_{-\textbf{k}}\Phi_{\textbf{k}}}.\nonumber
\end{eqnarray}
The corresponding electron wavefunction is $\Psi_e(\{{\bf
r}_i\})={\rm Pf}[g^{(l)}({\bf r}_i-{\bf
r}_j)]\prod_{i<j}(z_i-z_j)^{\tphi}e^{-{\sum_i|z_i|^2}/4}$. Here the
the pairing function $g^{(l)}_\textbf{k}$ is an eigenfunction of the
angular momentum
$L_z=\imth\hbar(k_y\partial_{k_x}-k_x\partial_{k_y})
=-\imth\hbar\partial_{\theta_\textbf{k}}$, i.e. $
g^{(l)}_\textbf{k}=e^{\imth l\theta_\textbf{k}}R_k$ with
$R_k=R(|\textbf{k}|)$ a \emph{real} function of $|\textbf{k}|$. The
parity of the pairing function must be \emph{odd} for spin-polarized
fermions, i.e. $g^{(l)}_{-\textbf{k}}=-g^{(l)}_{\textbf{k}}$; thus
$l$ must be an odd integer. The expectation value of an operator
$\hat O$ is given by $\langle\hat O\rangle=\langle G^{(l)}|\hat
O|G^{(l)}\rangle/\langle G^{(l)}|G^{(l)}\rangle.$

It is important to note that $\tilde g^{(l)}_{\textbf{k}}$ is not
independent of $g^{(l)}_{\textbf{k}}$ in the physical Hilbert space.
The hermiticity of electron pairing,
$$\langle\psi_e^\dagger(\textbf{r}+\textbf{x})
\psi_e^\dagger(\textbf{x})\rangle=\langle\psi_e(\textbf{x})
\psi_e(\textbf{r}+\textbf{x})\rangle^\ast$$ implies, through
relations (\ref{nutCF}), the following constraint between the two
pairing order parameters:
\begin{eqnarray}\label{hermiticity}
\langle\Pi(\textbf{r}+\textbf{x})\Pi(\textbf{x})\rangle \approx
(g_{\tphi}(r)\langle\Phi(\textbf{x})\Phi(\textbf{r}+\textbf{x})\rangle)^\ast
\end{eqnarray}
where $g_\tphi(r)$ is, remarkably, the {\em radial distribution
function} of the bosonic Laughlin wavefunction
$\Psi^{L}_{\tphi}(\{{\bf
r}_i\})\equiv\prod_{i<j}(z_i-z_j)^{\tphi}e^{-{\sum_i|z_i|^2}/4}$
which is shown in Fig.~\ref{fig1}(a) for $\tphi=2,4,6$. This
constraint describes mathematically how one NUCF feels the vortices
(correlation holes) around the others. Consequently, the two gap
functions $ \Delta_k=-\sum_{\bf k'}V^{(l)}({\bf
k,k}')\langle\Pi_{\bf -k'}\Pi_{\bf k'}\rangle$ and $\tilde
\Delta_k=-\sum_{\bf k'}\tilde V^{(l)}({\bf k,k}')\langle\Phi_{\bf
-k'}\Phi_{\bf k'}\rangle $ are related through the correlation
holes. We find,
\begin{eqnarray}
\tD{k}=\Delta_k+E_k\int_0^\infty
\frac{\Delta_{k^\prime}}{E_{k^\prime}}
\mathcal{H}_l(k,k^\prime)k^\prime\text{d}k^\prime \label{constraint}
\end{eqnarray}
where ${\cal H}_l(k,k^\prime)=\frac{1}{(2\pi)^2}\int
_0^{2\pi}d\theta_{kk^\prime}~e^{2\imth
l\theta_{kk^\prime}}h_\tphi({|{\bf k}-{\bf k}'|})$ and
$h_\tphi({|\bf q|})$ is the Fourier transform of the \emph{pair
correlation function} defined as $h_\tphi(r)\equiv g_\tphi(r)-1$.

Minimizing the ground state energy $E^{(l)}= \langle\hat H ^{\rm
CF}_{\rm eff}\rangle$ with respect to $R_k$ and $\tR k$, we obtain
the self-consistent Bogoliubov-de Gennes (B-dG) equations,
\begin{eqnarray}\label{B-dG:D,tD}
&R_k=\frac{E_k-\xi_k^\prime}{\tilde\Delta_k}=\frac{\Delta_k}{E_k+\xi_k^\prime},~~\tilde
R_k=\frac{E_k-\xi^\prime_k}{\Delta_k}=\frac{\tilde\Delta_k}{E_k+\xi_k^\prime}
\end{eqnarray}
where $E_k=\sqrt{(\xi_k^\prime)^2+\Delta_k\tilde\Delta_k}$ is the
quasiparticle dispersion and $\xi^\prime_k=\xi_k+\xi^{PH}_k$ is the
renormalized dispersion due to the Coulomb interaction in the
particle-hole channel where
$\xi_k^{PH}=-\frac{2k_F\lambda}{m_b}\int_0^\infty
n_{k^\prime}|k^\prime|\text{d}|k^\prime|M^{(0)}_{|k|,|k^\prime|}$
and $M^{(0)}_{|k|,|k^\prime|}$ is given by
${|k-k^\prime|}^{-1}=\sum_{l}M^{(l)}_{|k|,|k^\prime|}e^{\imth
l\theta}$. The dimensionless interacting strength $\lambda$ measures
the ratio of the Coulomb interaction strength to the Fermi energy:
$\lambda=V_c(k_F^{-1})/4\epsilon_F=
\frac{e^2m_b}{8\pi\epsilon_0\epsilon_rk_F}$
with $k_F=\sqrt{4\pi n_e}=\sqrt{2\nu}/\ell_B$ the Fermi wavevector.
The momentum distribution is $n_k=(E_k-\xi_k^\prime)/{2E_k}$.

It is important to stress that the relation (\ref{constraint}) between the two canonical conjugate gap functions projects the non-unitary theory onto the physical Hilbert space. Assuming $\bar\Delta_k=\Delta_k^*$ \cite{japan} would violate this constraint and fail to capture the correlation-hole effects. One can show that, if $\bar\Delta_k=\Delta_k^*$ is assumed, the solution of the BdG equation
(\ref{B-dG:D,tD}) is $\bar\Delta_k=\Delta_k^*=0$ for all $k$.

We solved the BdG equations (\ref{B-dG:D,tD}) under the constraint
(\ref{constraint}) for possible values of $l$ and found that the
leading pairing instability has indeed $l=-1$, i.e. $p_x-\imth p_y$
wave symmetry, which will be our focus. For weak Coulomb
interactions $\lambda<\lambda_{c1}$, the two conjugate gap functions
are in-phase, i.e. $\Delta_k\tD{k}>0$ and $R_k\tilde R_k>0$ for all
$k$, as shown in Fig.~\ref{fig1}(b). In this case, the asymptotic
solutions can be obtained analytically in the long wavelength limit:
$\Delta_k,\tD{k}\propto\vert k\vert$ as $\vert k\vert\to0$. Thus,
the pairing function $g^{(-1)}_k\propto 1/k$ and the paired state is
asymptotically a MRP. To study the paired state quantitatively for
all $k$, we numerically solve for the gap functions using an
ultraviolet cutoff in momentum space, {\it e.g.}, $|k|\leq
\Lambda=1.4 k_F$. It is clear from Fig.~\ref{fig1}(b) that $1/R_k$
deviates significantly from the linear behavior such that the
wavefunction of the paired state ${\rm Pf}[g^{(-1)}({\bf r}_i-{\bf
r}_j)]\Psi^L_\tphi(\{{\bf r}_i\})$ is different from a MRP at
shorter distances $|{\bf r}_i-{\bf r}_j|\leq\ell_B$.

\setlength{\unitlength}{1in}
\begin{figure}
\begin{picture}(3.7,3.1)
\put(-0.3,0){\makebox(3.7,3.1){\includegraphics[width=3.7in]{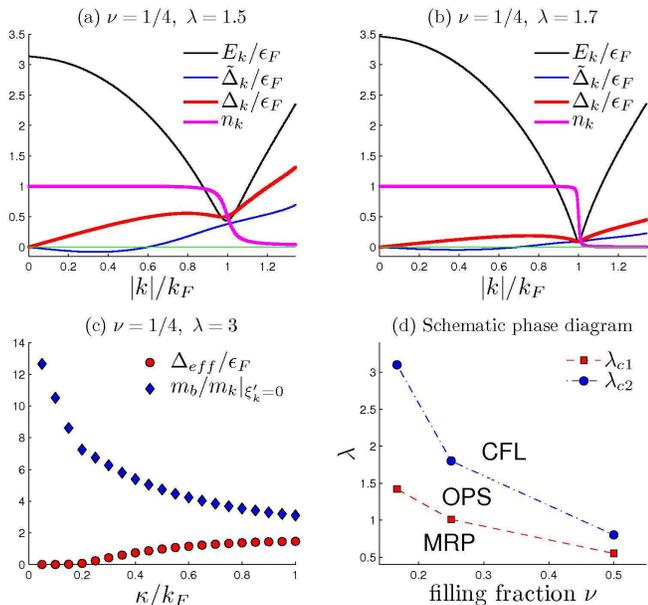}}}
\end{picture}
\caption{(a) and (b): The evolution of the quasiparticle dispersion
$E_k$, the gap functions $(\Delta_k,\tD{k})$, and the momentum
distribution $n_k$ at $\nu=1/4$ with increasing $\lambda$. The
excitation gap closes at $\lambda_{c2}\approx1.8$. (c): The effects
of screening at fixed $\lambda=3$. The excitation gap $\Delta_{\rm
eff}$ and the effective mass $m_k$ at the Fermi level
($\xi_k^\prime=0$) are plotted as a function of the inverse
screening length $\kappa$. (d): The phase diagram of the
even-denominator FQHE at $\nu=1/2,~1/4,~1/6$. } \label{fig2}
\end{figure}

Remarkably, a completely new paired state emerges for intermediate
Coulomb interactions. When $\lambda>\lambda_{c1}$, the gap function
$\tilde\Delta_k$ changes sign at $k_0$ where $R_k$ is singular as
shown in Fig.~\ref{fig1}(c) and Fig.~\ref{fig2}(a,b). This
singularity causes the pairing function $g^{(-1)}({\bf r})$ to
oscillate in real space as shown in Fig.~\ref{fig1}(d). We find that
in the long-wavelength limit its amplitude decays as $1/\sqrt{r}$
according to $g^{(-1)}({\bf
r})\sim(\sqrt{|z|}/z)\cos(k_0|z|-\frac34\pi)$. Despite the sign
change in the gap function $\tilde\Delta_k$, this oscillatory
pfaffian state (OPS) remains fully gapped and topologically stable
with quasiparticle dispersions shown in
Figs.~\ref{fig2}(a,b). The topological winding number associated with the
mapping, via the pairing function $g_k$, from a compactified complex
plane $k\equiv k_x+\imth k_y$ to the two-sphere $\hat
n_k\equiv(2~{\rm Re} g_k,2~{\rm Im} g_k,1-|g_k|^2)/(1+|g_k|^2)$ is
{\it one}, generic of a chiral
$p$-wave paired state. This state can be detected by
spectroscopy measurements since the singularity at $k_0$ produces a
kink in the quasiparticle density of states that moves toward the
Fermi level as $k_0$ approaches $k_F$ with increasing $\lambda$.
For $\lambda >\lambda_{c2}$, the paired state is destroyed and the
Rezayi-Read CFL becomes the variational ground state. This
quantum phase transition is signaled by the closing of the
quasiparticle excitation gap $\Delta_{\rm eff}\equiv{\rm min}
_k\{E_k\}$ shown in Fig.~\ref{fig2}(a,b) near the transition. We
find that both $\lambda_{c1}$ and $\lambda_{c2}$ increase
monotonically with $\tilde\phi=\nu^{-1}$, as shown in
Fig.~\ref{fig2}(d).

Since the interaction at short distances is reduced efficiently by
the correlation-hole, the stability of the CFL against pairing must
rely on the long-range part of the Coulomb interaction. As a result,
the effects of screening become important. To illustrate this, we
consider the 3D screened Coulomb interaction of the Yukawa form:
$V_{\rm sc}(r)=V_c(r)e^{-\kappa r}$, where $\kappa$ is the inverse
screening length. Fig.~\ref{fig2}(c) shows that a CFL stabilized by
a large enough $\lambda>\lambda_{c2}$
can be driven through a continuous transition into a paired state
by increasing screening, i.e. reducing the screening length.
Concomitantly, the logarithmic divergence of the effective mass in
the CFL at $\kappa=0$ \cite{read} is removed. We note that the MRP has been shown to be more stable when finite layer thickness is considered in the Coulomb interaction at $\nu=5/2$ \cite{sankar}.

To summarize, we have shown that the correlation hole, i.e. the
binding of electrons to quantized finite-size vortices, is crucial
for forming the paired quantum Hall states.
The effective interaction is gauge-field free, instantaneous, and
favors chiral $p$-wave pairing. The pairing potential is a direct
consequence of the lowering of the Coulomb energy due to the
correlation hole.
~We find that, with increasing Coulomb interaction strength, the
ground state evolves from a $p-ip$ state asymptotically equivalent
to the MRP, to a new oscillatory pfaffian state and finally to a CFL
via a continuous phase transition as illustrated in
Fig.~\ref{fig2}(d).

Recently, FQHEs at $\nu=1/2$ and $1/4$ have been observed in wide
GaAs quantum wells\cite{exp1,exp2,exp3} with higher electron density
than in previous experiments that reported no signs of FQHE.
Indeed, there are direct evidences
\cite{exp1,exp2} that the signatures of FQHE become stronger with
increasing electron density. This is consistent with our theory
since $\lambda$ is proportional to $k_F^{-1}\sim n_e^{-1/2}$; a
smaller density would tend to destabilize the paired state. 
Hence, the paired quantum Hall states proposed in this work can be
prospective candidates for the observed even-denominator FQHE in the
LLL.

Y.-M. Lu and Z. Wang thank ITP, CAS, and KITP at UCSB for hospitality.
This work was supported by NNSF and the national program for basic
research of MOST of China, a fund from CAS (YY), and by DOE Grant
DE-FG02-99ER45747 (YML,ZW) and NSF PHY94-07194.

\end{document}